\documentclass[times,twocolumn,final,authoryear]{elsarticle}

\usepackage{jasr}
\usepackage{framed,multirow}

\usepackage{amssymb}
\usepackage{latexsym}
\usepackage[T1]{fontenc}
\usepackage[utf8]{inputenc}
\usepackage{makecell}
\usepackage{lscape}

\usepackage{url}
\usepackage{xcolor}
\definecolor{newcolor}{rgb}{.8,.349,.1}

\usepackage[citebordercolor=white]{hyperref}

\newcommand{\apjl}{ApJ Let.}
\newcommand{\aj}{AJ}

\newcommand{\aap}{A\&A}

\newcommand{\mnras}{MNRAS}
\newcommand{\memsai}{Mem. Societa Astronomica Italiana}

\newcommand{\nat}{Nature}

\newcommand{\pasa}{Publications of the Astronomical Society of Australia}
\newcommand{\planss}{Planetary and Space Science}

\newcommand{\nar}{New Astronomy Reviews}

\journal{Advances in Space Research}

\begin{document}

\verso{Domingos Barbosa \textit{etal}}

\begin{frontmatter}

\title{Radio astronomy and Space science in Azores: enhancing the Atlantic VLBI infrastructure cluster}%

\author[1]{Domingos Barbosa}
\ead{dbarbosa@av.it.pt}
\author[1]{Bruno Coelho}
\author[2,1]{Sonia Ant\'on}
\author[1,8]{Miguel Bergano}
\author[1,7]{Tjarda Boekholt}
\author[3]{Alexandre C.M. Correia}
\author[4]{Dalmiro Maia}
\author[1,5]{Jo\~ao Pandeirada}
\author[1,5]{Val\'erio Ribeiro}
\author[5,6]{Jason Adams}
\author[1,5]{Jo\~ao Paulo Barraca}
\author[1,5]{Diogo Gomes}
\author[4]{Bruno Morgado}

\address[1]{Instituto de Telecomunicações, 3810-193 Aveiro, Portugal}
\address[2]{CIDMA, Campus de Santiago, 3810-183 Aveiro, Portugal}
\address[3]{ CFisUC, Departamento de F\'isica, Universidade de Coimbra, 3004-516 Coimbra, Portugal}
\address[4]{CICGE, Faculdade de Ciências da Universidade do Porto, 4169-007 Porto, Portugal}
\address[5]{Universidade de Aveiro, 3810-193 Aveiro, Portugal}
\address[6]{AMSL Holding, Veldhoven, The Netherlands}
\address[7]{Rudolf Peierls Centre for Theoretical Physics, Parks Road, Oxford OX1 3PU, United Kingdom}
\address[8]{ESTGA - Universidade de Aveiro, 3750-127 Aveiro, Portugal}
\received{5 March 2021}
\finalform{20 May 2021}
\accepted{25 May 2021}
\availableonline{1 June 2021}
\communicated{Mr. X}

\begin{abstract}
Radio astronomy and Space Infrastructures in the Azores have a great scientific and industrial interest because they benefit from a unique geographical location in the middle of the North Atlantic allowing a vast improvement in the sky coverage. This fact obviously has a very high added value for: i) the establishment of space tracking and communications networks for the emergent global small satellite fleets ii) it is invaluable to connect the radio astronomy infrastructure networks in Africa, Europe and America continents using Very Large Baseline Interferometry (VLBI) techniques, iii) it allows excellent potential for monitoring space debris and Near Earth Objects (NEOs).  There is in S. Miguel island a 32-metre SATCOM antenna that could be integrated in advanced VLBI networks and be capable of additional Deep Space Network ground support. This paper explores the space science opportunities offered by the upgrade of the S. Miguel 32-metre SATCOM antenna into a world-class infrastructure for radio astronomy and space exploration: it would enable a Deep Space Network mode and would constitute a key space facility for data production, promoting local digital infrastructure investments and the testing of cutting-edge information technologies. Its Atlantic location also enables improvements in angular resolution, provides many baseline in East-West and North-South directions connecting the emergent VLBI stations in America to Europe and Africa VLBI arrays therefore contributing for greater array imaging capabilities especially for sources or well studied fields close to or below the celestial equator, where ESO facilities, ALMA, SKA and its precursors do or will operate and observe in the coming decades.
\end{abstract}


\end{frontmatter}


\section{Introduction}
\label{S:1}

Azores has a unique location in the Atlantic spreading over 600 km, the archipelago has islands dispersed across three tectonic plates: North American, Eurasian and African plates. In fact, the plate junctions pass through its central group (Islands of Faial, Pico, S. Jorge, Terceira and Graciosa). In the 1990s, NASA used the Azores as an important geodetic satellite observation site to improve the spatial navigation with GPS \citep{1993Bryant}. Furthermore, since the early 2000s there has been considerable interest for a station in the Azores for science and innovation to provide a unique radio astronomical, space sciences and navigation facility.

Azores is now developing a new infrastructure: Atlantic International Research Center (AIR), which is an international platform integrating research and innovation on multi-related areas such as climate change, earth observation, energy, space and oceans. The Florianopolis Declaration \citep{florianopolis} led to the confirmation of this new Atlantic interaction level \citep{florianopolis2,florianopolis3} which will aggregate complementary aspects of a new critical infrastructure towards high impact space sciences.

Radio astronomy and space sciences are areas that can take advantage of the unique location of Azores, and projects on either the installation of new stations in St. Maria or Flores islands or the update and retrofit of a large space communications (SATCOM) dish in S. Miguel, have been proposed. One of the projects that has already began is the deployment and operation of 13.2 metre radio telescope with Very Long Baseline Interferometry (VLBI) capabilities of the Rede Atlântica de Estações Geo-Espaciais (RAEGE, or Atlantic Network of Geo-Spatial Stations). This telescope is deployed at St. Maria island (Azores Western group), a project led by the National Geographical Institute of Spain (IGN-Spain) and the Government of the Autonomous Region of Azores \citep{2015Hass}.

On the other hand, this is also the era of great developments in the South Atlantic: (a) the Square Kilometre Array (SKA) will become the largest distributed scientific facility of the XXIst century, will yield new and unique insights, from the quest for the Origins to the existence of life elsewhere, to other aspects like the matter and energy content of the Universe and our cosmic history; (b) the African VLBI Network (AVN), a network of radio telescopes throughout Africa, mostly from upgrades of SATCOM Intelsat Standard-A 32-m antennas, that will have the capability of extending the VLBI worldwide network. SKA and AVN are also expected to provide during their lifetime important transformational telemetry support and data downlink to the new generation of Deep space probes \citep{2016Schutte}. In this paper we aim to address the potential of Azores location for multi-disciplinary research, in the framework of existing infrastructure (RAEGE dishes, putative updated SATCOM dish, SATCOM for Cubesats, AIR) and new high sensitivity facilities like SKA, AVN and European VLBI Network (EVN).

\subsection{From Space Tracking to Space Ports}

Recent developments in Azores target the emergent global small satellite markets that are rapidly growing, especially in the nano and micro-satellites mass ranges. The rapidly decreasing costs of such small satellites coupled to the ever increasing capabilities enabled by many COTS developments and availability of state-of-art technologies make the new generation of satellites more efficient than previous more massive generations at a fraction of their cost. As such, small satellites constellations or swarms have become an interesting concept enabling innovative missions that were not previously possible, and which require innovative ground support infrastructures. These range from Low Earth Objects (LEO) cubesats for Earth observations (climate change, ocean monitoring, etc) to science missions like the Chinese Chang'e-4 mission to the far side of the Moon that will include a pair of microsatellites to be placed in lunar orbit to test low frequency radio astronomy and space-based interferometry \citep{2017Ye}. 
Notably, small satellites” cubesats swarms in LEO orbits are commercially and “environmentally” sustainable: high atmosphere friction drag self-cleans constellations with time and they do not contribute for the long term space debris field affecting higher orbits (they simply fall down and disintegrate with time). Their number is also growing very fast and the number of expanding large constellations does indeed represent a data challenge in terms of the aggregated data transmission and in orbit occupancy. 
Typically, smaller and fast tracking stations are the prime choice for low frequency-band communications. However, these smaller stations do not possess the sensitivity for ground support to missions requiring Deep Space Network capabilities because these missions are very faraway or they rely on communications using higher frequency bands (from C, X-band or even Ka bands). Deep-space missions largely rely on the use of radio tracking for their orbit determination and the associated parameter estimation, using Doppler data (closed loop) obtained by, for example, NASA’s Deep Space Network (DSN) and ESA’s TRACKing station network (ESTRACK). However, VLBI measurements using the Planetary Radio Interferometry and Doppler Experiment (PRIDE) technique (open loop) have been used on a large number of past and current planetary missions to provide unsurpassed precision about spacecraft accurate angular positions in the sky, i.e. spacecraft lateral position (right ascension 
 $\alpha$ and declination $\delta$) measurements with an uncertainty of approximately 1.0 nrad ($\sim$200 $\mu$arcsec) or about 50 metre precision at a distance of 1.4 AU. PRIDE techniques are used by the large radio telescopes from the European VLBI Network (EVN, EurAsia) and the Very Large Baseline Array (VLBA, USA) in collaboration with major space agencies.
PRIDE does not require special capabilities from the mission’s on-board instrumentation and it can be applied to almost any radio signal from a spacecraft, provided the spacecraft signal is minimally powerful and phase-stable. In fact, most of Deep space planetary probes can be considered to be PRIDE or VLBI customers.

PRIDE/VLBI  has been used as a multi-purpose, multi-disciplinary enhancement of planetary missions science return on a large number of past and current planetary missions, to name a few : VEGA Venus atmosphere balloons, Ulysses solar orbiter \citep{1996Folkner}, the Huygens Probe during its descent to the surface of Saturn’s moon Titan, from the Cassini-Huygens mission and the VLBI tracking of the Cassini spacecraft at Saturn \citep{2004Pogrebenko,2005Lebreton,2015Jones}, Chang'E-1 flight to the Moon \citep{2010Jianguo}, VLBI tracking with the European VLBI Network (EVN) antennas of  the controlled impact of ESA’s Smart-1 probe on the surface of the Moon \citep{2006Avruch},  VLBI tracking of NASA’s Mars Exploration Rover B spacecraft during its final cruise phase \citep{2007Lanyi}, VLBI tracking of the solar sail mission IKAROS \citep{2011Takeuchi}, ESA’s Venus EXpress (VEX) VLBI spacecraft observations \citep{2012Duev}, the ESA Mars Express (MEX) Phobos-flyby \citep{2015Park,2016Duev} monitored by the EVN radio telescopes and the SKA precursor Murchinson Radio Observatory (MRO), and NASA’s Juno flyby of Jupiter \citep{2015Jones}.

\section{The emergence of science and space radio Infrastructures in Azores}

In the last decade, Azores has seen the development of new infrastructures using radioastronomical techniques (mainly for geodesy) and the deployment of new space stations for space telemetry and services or science ground segment that complement the local space SATCOM cluster much necessary to guarantee communication redundancy between the most distant Azores islands and the biggest island of S\~ao Miguel. The RAEGE network will be composed of four 13.2 m Geodetic Fundamental Stations in Spain (Yebes and Canary Islands), and Portugal (Azores Islands of Santa Maria and Flores) as part of the developments required to set up a VLBI Geodetic Observing System, VGOS \citep{2010gomez, 2015gomez}. VGOS became the standard VLBI system for geodesy and astrometry \citep{2012Petrachenko}, being part of the Global Geodetic Observing System (GGOS) of the International Association of Geodesy (IAG), which integrates different geodetic techniques to provide the geodetic infrastructure necessary for monitoring the Earth system and for global change research.
Currently, there are in full operations the existing stations in Yebes (near Madrid, European tectonic plate) and in St. Maria (African tectonic plate), both operated by IGN. Besides this, a second Azores RAEGE radio telescope is planned for deployment in Flores Island (North American tectonic plate) by 2022-2024. The Canarias Islands RAEGE station (African tectonic plate) is located in Cruz de Acusa, at an altitude of 1100 m, in the island of Gran Canaria (Canary Islands) and is in construction since 2019. The Canarias station will be equipped with a new generation wide-band receiver (2-14 GHz) built in Gran Canaria, in cooperation with the University of Las Palmas de Gran Canaria (ULPGC-IDETIC). Figure 1 shows the RAEGE network.  In full operation, these four geodetic VLBI stations will add an improved capability in global changes studies and monitoring of tectonic movements of a wide and very dynamic geophysical region working together as a radio "Iberoscope" with a maximum baseline of about 2200 km. The RAEGE telescopes have a very stringent optical design, optimized for geodetic VLBI observations enabling the measurement of the Earth reference frame with 1 mm accuracy and are designed to enable observations up to or above Ka band (40 GHz). Due to their size optimized for geodesy applications, the RAEGE telescopes lack the capability for very high resolution astronomy and deep space experiments with planetary missions to the outer Solar System. Also, the high humidity atmospheric content resulting from the combination of the central-Atlantic climate and the low altitude of these facilities make observations above Ka band very difficult. 

\begin{figure}[h]
\centering\includegraphics[width=1\linewidth]{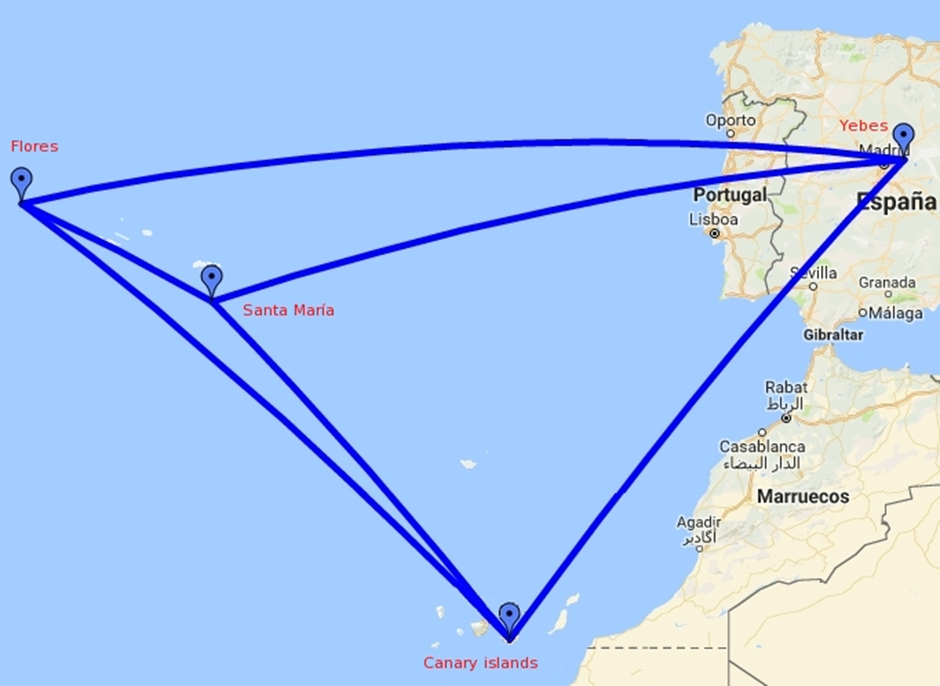}
\caption{The RAEGE network, covering most of Macaronesia, or the central and North-West Atlantic Ocean. When fully deployed, RAEGE will have VLBI stations spread over the American, European and African plates. (Credit: RAEGE Project)}
\end{figure}

Recently, the Portuguese Science and Technology Foundation (FCT) and the Portuguese Space Agency (PTSpace) have deployed a 15-metre station for space service in Azores St. Maria Island, (coming from the ESA DSN complex near Perth, Australia) as part of the ground segment of Proba-3 solar physics mission. Besides this science service, this station has been considered  to greatly enhance downlink coverage in the Atlantic for Low Earth Orbit (LEO) satellites, in particular for ESA support and the growing LEO constellations. However, support to Deep Space missions to very distant targets in the Solar system may require larger antennas, of the class 25-32 metres, that were dedicated previously to SATCOM activities, and are currently redundant. Fortunately, Azores also possesses a decommissioned 32-metre Intelsat Standard-A SATCOM in S. Miguel island, an Earth station from former Marconi/PT Comunicações, currently from Altice Portugal and available to retrofit (see Figure 2). This Earth station, after retrofitting, can cover the Deep Space segment and open Azores to the high impact contributions of Deep Space missions, enhancing the AIR footprint to potential collaborations with major space agencies (NASA, ESA) on deep solar system space missions. 

More recently, a VHF ground station, St. Maria-PRT1, was installed by the RAEGE station as a node of the ground alert system of the space mission SVOM (Space-based multi-band astronomical Variable Objects Monitor), a Franco-Chinese mission dedicated to the study of the most distant explosions of stars, the gamma-ray bursts \citep{2020ExA....50...91D,2018MmSAI..89..266C,2016arXiv161006892W}. SVOM is to be launched at the end of 2021 by the Chinese Long March 2C rocket from the Xichang launch base. See Table 1 for a more detailed description.

\begin{figure*}[h]
\centering\includegraphics[width=1\linewidth]{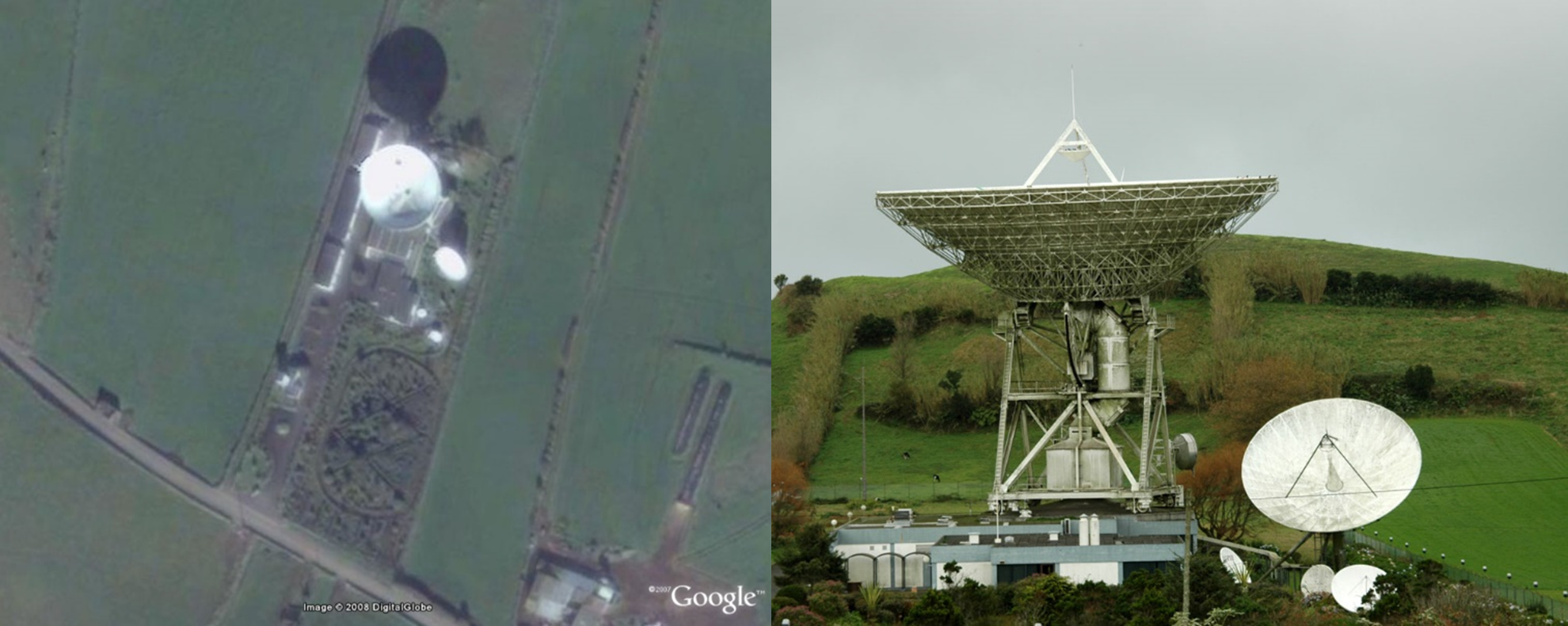}
\caption{SATCOM Earth Station in São Miguel (Azores) – Intelsat Standard A 32-metre parabolic antenna, Altice Portugal. We can see the waveguide system below the dish.}
\end{figure*}

\begin{table*}[h]
\centering
\begin{tabular}{l c c l c }
\hline
\textbf{Station} & \textbf{\makecell{Dish Size\\(m)}} & \textbf{\makecell{Freq.\\(GHz)}} & \makecell{\textbf{Lat.(deg)}\\ \textbf{Long.(deg)}} & \textbf{Capabilities}\\
\hline
\hline
\makecell{RAEGE \\St. Maria\\(in operations)} & 13.6 & 2-40 & \makecell{36.98518 \\ -25.12588} & \makecell{Geodetic VLBI,\\ Geodesy,\\ global change studies,\\ GNSS}\\
\hline
\makecell{RAEGE \\Flores\\(in preparation)} & 13.6 & 2-40 & \makecell{$\sim{39.45}$ \\ $\sim{-32.2}$} & \makecell{Geodetic VLBI,\\ Geodesy,\\ global change studies}\\
\hline
\makecell{SVOM\\St. Maria-PRT1 station\\(deployed, circa RAEGE)} &-& 0.137-0.138 (VHF) & \makecell{36.9858\\-25.1262} & \makecell{SVOM space mission \\ Ground segment \\ GRB / transient sky Alert network}\\
\hline
\makecell{St. Maria\\ESA Tracking\\(in operations, EDISOFT)} & 5.5 & 2.3, 5, 7-8.5 & \makecell{36.99725\\-25.13583} & \makecell{Kourou launch\\ tracking, CleanSeaNet,\\ satellite detection\\of oil slicks}\\
\hline
\makecell{VLBI SATCOM\\Station S. Miguel\\(proposed)} & 32 & 5-10 & \makecell{37.79088\\-25.66473} & \makecell{VLBI, Very high\\resolution Astronomy\\+Space VLBI; Deep\\space network doppler\\tracking; Space debris}\\
\hline
\makecell{PTSPACE \\Ground Station (deployed)} & 15 & 2.3, 7.1-8.5 & \makecell{36.99651\\-25.13.6514}  & \makecell{Proba-3 DSN;\\LEO polar + helio\\synchrounous orbits$^{*}$}\\
\hline
\makecell{LEO Station\\(planned)} & $2.4/4.5$ & $\sim2.3-4$, $\sim7-10$  & St. Maria Island & \makecell{LEO tracking; Rx-Tz\\GGSN for cubesat,\\ formation flying$^{\dagger}$}\\
\hline
\end{tabular}
\label{tab1}
\caption{Summary of existing and planned Space and radio astronomy infrastructure in Azores.; $^{*}$Based on FCT/Edisoft-Thales.; $^{\dagger}$based on Heliaq information.}
\end{table*}

\section{From SATCOM to VLBI Networks}


Satellite communications (SATCOM) were introduced as a carrier of telephone, data and TV signals, in order to supplement undersea cables and provide connectivity to peripheral regions. The radio frequency bands allocated for this service are essentially those within the frequency range known as C-band, i.e. 5.925 to 6.425 GHz for uplink and 3.700 to 4.200 GHz for downlink. To provide support to these services, Intelsat Standard-A antenna Earth stations with large dishes, typically about 32 metres in diameter, were designed and deployed around the world, in particular in the 70s and 80s years. From mid-80s the new satellite communication technologies and the ever increasing coverage and bandwidth of modern digital ground optical fiber networks paved the way for much smaller ground station apparatus and the largest stations became redundant.

These Intelsat dishes share a similar design and focal length to radio telescopes. This means that they are relatively cheap and easy to convert into scientific or space service facilities. Not surprisingly, there are a number of good examples across the world of transforming redundant Intelsat Standard A dishes into scientific world class facilities: the 30-m dish at Ceduna in Australia \citep{2005McCulloch} was transformed into a radio telescope in 1985 by the University of Tasmania, in Atlanta (USA), a 30-m SATCOM antenna, was acquired and fully renovated as radio telescope by the Georgia Institute of Technology \citep{1999DeBoer}; in Japan, near Yamaguchi, a 32-m antenna \citep{2002Fujisawa} was upgraded to science operations by the National Astronomical Observatory of Japan (NAOJ); a SATCOM transformation in Elfordstown near Cork in the Republic of Ireland \citep{2005Gabuzda} was considered to enhance the resolution and uv-coverage of both the Multi-Element Radio-Link Interferometer Network (e-MERLIN) and the European Very Long Baseline Interferometer (VLBI) Network (EVN). Additionally, in Latvia, the Ventspils International Radio Astronomy Center (VIRAC) has recovered a 32-metre old Soviet station to an operating radio telescope, a pivotal facility of the Baltic research in space science with a particular focus in planetary sciences (comets, NEOs, etc)\citep{2013BaltA,2013BaltB}.

With the SKA Phase1 Mid Frequency telescope to start construction in South Africa, there is now a considerable interest in radio astronomy across the African continent after the launch of the Africa VLBI Network (AVN). AVN is in the process of the constitution of radio astronomical and space science communities across Africa from leveraging on the conversion of former SATCOM in Africa. In fact, there are about 29 documented 30-m class telecommunications antennas in 19 African countries though some no longer exist. Ghana just saw the first successful African example of an Earth station in Kutunse transformed into a world class radio astronomical and space facility \citep{2011Gaylard,2012Nordling,2012Perks}, with Zambia and Kenya to follow.

Another very interesting example is the transformation of the famous and former British Telecom SATCOM facilities at the Goonhilly Earth Station (GES) in Cornwall, UK \citep{2011Heywood}. GES, currently a private company, is now ready to perform radio astronomy observations with a conversion of a 25-metre dish to radio telescope, Deep Space tracking and provides ground support servicing to the UK Space Port Cornwall activities in Newquay, Southwest England; this includes space navigation and telemetry support to near future cubesat constellations in LEO orbits and support to private space operators like Virgin Orbit. Besides this conversion, two of the Goonhilly antennas, Goonhilly-1 (GHY-1, or "Arthur") and Goonhilly-3 (GHY-3, or "Guinevere") are also planned to be included in radio interferometry arrays (e-MERLIN and EVN) as part of their conversion to radio astronomy facilities\citep{2011Heywood,2011arXiv1103.3600K}. The later GHY-3 may be suitable to use for higher-frequency (especially K-band) observations. The addition of several Goonhilly antennas is expected to bring an enhanced array performance and consequently impact the scientific return of the e-MERLIN and  EVN Legacy observing programs.

\subsection{Analysis of the Azores contribution for different VLBI networks. }
It is well known the angular resolution of an interferometer is determined by the longest "projected" baseline  $\propto\lambda/b$, where $\lambda$ is the observation wavelength and $b$ is the baseline. Typically the Azores antenna will have baselines of about 1900 km to the nearest VLBI stations (Robledo and Yebes, in Spain), $\sim6000$ km to Mexico and spanning up to 11000 km towards the VLBI stations in South Africa or Eastern Asia (China and Korea). As an example, for observations carried at 5GHz (6cm wavelength), the addition of the Azores station to EVN (Europe; see Fig. 3) would improve typical angular resolution from 2.1 miliarcseconds ({\it mas}) to $\sim$~1.6 {\it mas} and if we add it to the longest continental baseline EVN (with Asia), the angular resolution would improve from 1.54 {\it mas} to $\sim$~1.28 {\it mas}. Considering the baseline connections to other networks (EVN-Africa and EVN-Mexico) the Azores antenna will contribute to cover the so called "Atlantic gap" with intermediate baselines therefore improving the overall image fidelity and enabling a better cross-correlation studies with ALMA or the core of SKA Phase1, compact arrays that are planed for good imaging capabilities. The contribution of a single VLBI radio telescope to existing networks can be better perceived by so called the UV-plane, i.e. the Fast Fourier Transform of image sky brightness. The better we fill-in the UV-plane, the higher the quality and fidelity of the images of the sources (astronomical objects, spacecrafts). Figures 3, 4, 5, 6 and 7 demonstrate the improvements in the UV coverage of adding the contribution of the Earth Station's 32-metre antenna in São Miguel (Azores) to different configurations of VLBI stations, they also include a representation of the corresponding baselines and station's locations. The simulated UV-plane coverages correspond to an Earth rotation period (24h), considering all the stations that observe the source QSO 0234+285 with elevation $\geqslant$15º at 5GHz (C-band). The simulations used a set of python tools derived from \citep{2000tra..book.....R} and considered for the moment a single channel centred at 5GHz. Once the C-band receiver properties (bandwidth and number of channels) become established according to VLBI standards, we will perform a follow up analysis using CASA Synthesis \& Single Dish Reduction package \citep{2007ASPC..376..127M}  sm tool. The addition of a new antenna into an existing VLBI array provides enhanced imaging sensitivity and in the case of Azores, this will translate in the enhancement of the spatial sensitivity by adding many new baselines that traces new sets of spatial scales. For simplicity, we consider here the addition into a homogeneous network where the the imaging sensitivity ${\Delta I}$ in Jy becomes described by \citep{1989ASPC....6..355W,1999ASPC..180..171W,2011arXiv1103.3600K} : 
${\Delta I} = (\eta_{b} SEFD) / \sqrt{N\,(N - 1) \, N_{\rm Stokes} \Delta \nu \Delta t}$, where N is the number of antennae, SEFD is the system equivalent flux density, which is defined by SEFD = $T_i /  K_i$, where $T_i$ is the system temperature of antenna $i$ respectively in Kelvin and $K_i$ is the antenna sensitivity, $N_{\rm Stokes}$ is the number of polarisation products and  $\eta^2_{b}$ is an efficiency term accounting for
losses due to digitisation of the signal for correlation. If the considered antennas in Table 2 were modeled as an homogeneous array, the addition of the Azores antenna would improve array image sensitivity by 7\% (EVN alone) and 8\% (for EVN+AVN+TM65). 

In Fig. 3 is shown the UV-plane for Azores in addition to a current EVN configuration, Fig. 4 shows the same stations with addiction of an upcoming AVN configuration, and in Fig. 5 is added one more station corresponding to Tianma65 in China. The addition of an Azores dish alone contributes to (a much needed) East-West baseline relative to the main stations in Eurasia and Africa (which provide a North-South baseline). Long baselines detect the most compact, otherwise unresolved structures. This remarkable information highlights the great scientific potential of such a contribution. In Figure 6 is shown the UV-plane for an Azores plus an AVN configuration, with addition of SKA-mid contribution, and finally Figure 7 shows the UV-plane considering Azores, the previous mentioned EVN configuration with addition of Tulancingo antenna in Mexico, in this last case Azores contribution is most relevant to fill-in the UV coverage, providing intermediate baselines.
The baselines, and station's locations are summarized in Table 2. 

\begin{figure*}[h]
\begin{tabular}{ll}
\centering\includegraphics[width=0.475\linewidth]{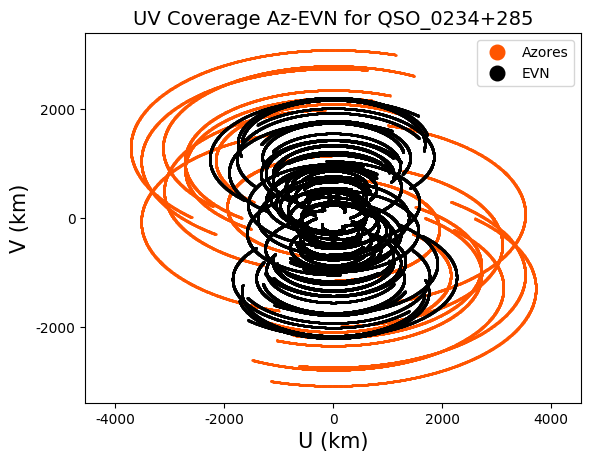}
&
\centering\includegraphics[width=0.475\linewidth]{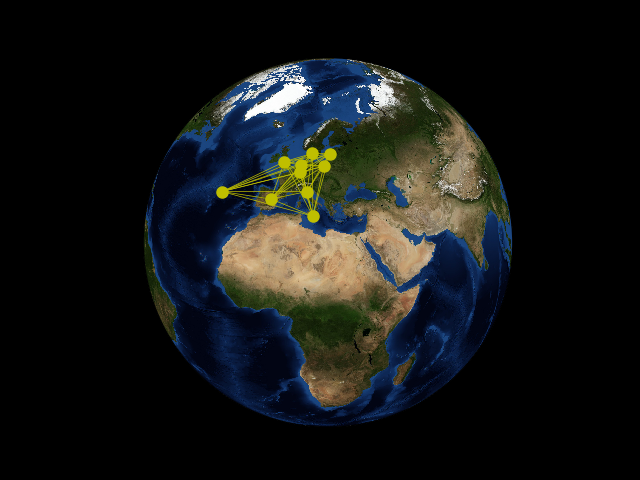}
\end{tabular}
\caption{Left: UV-plane coverage of European VLBI Network stations + Azores (in red). Azores provides the Atlantic unique and fantastic coverage and enhances the dynamic range of observations. Right: Baselines of European VLBI Network stations + Azores used in the simulation. In this case Azores would provide the largest baseline, contributing to improve the resolution of the European Network. Source used: QSO 0234+285.}
\end{figure*}

\begin{figure*}[h]
\begin{tabular}{ll}
\centering\includegraphics[width=0.475\linewidth]{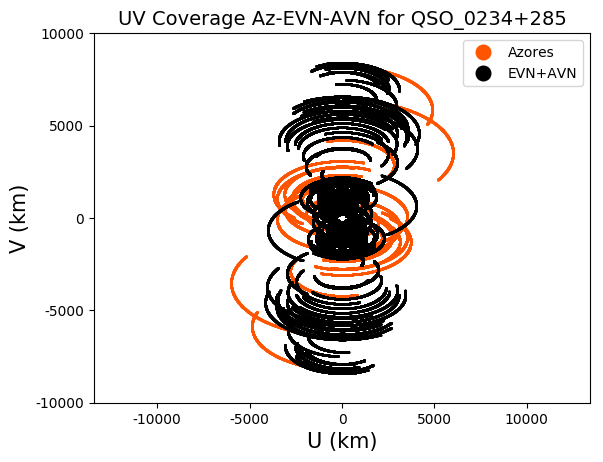}
&
\centering\includegraphics[width=0.475\linewidth]{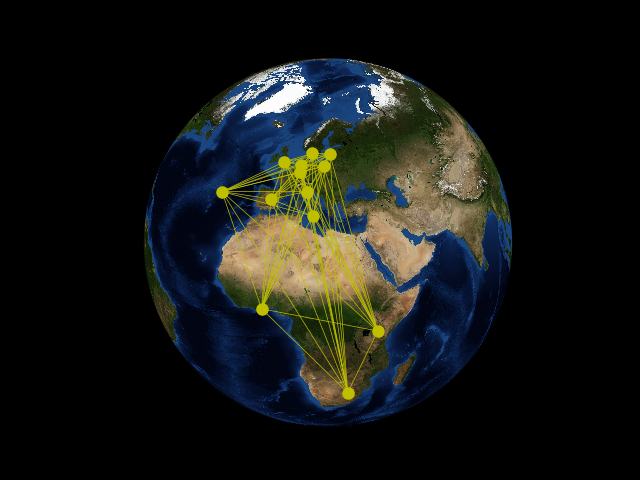}
\end{tabular}
\caption{Left: UV coverage contributions of European VLBI Network + Africa VLBI Network (Ghana – Kuntunse; Longonot – Kenya; HART – South Africa) + Azores (in red). Azores Atlantic coverage is unique. Right: Baselines of European VLBI Network + Africa VLBI Network (Ghana – Kuntunse; Longonot – Kenya; HART – South Africa) + Azores used in the simulation. Source used: QSO 0234+285.}
\end{figure*}

\begin{figure*}[h]
\begin{tabular}{ll}
\centering\includegraphics[width=0.475\linewidth]{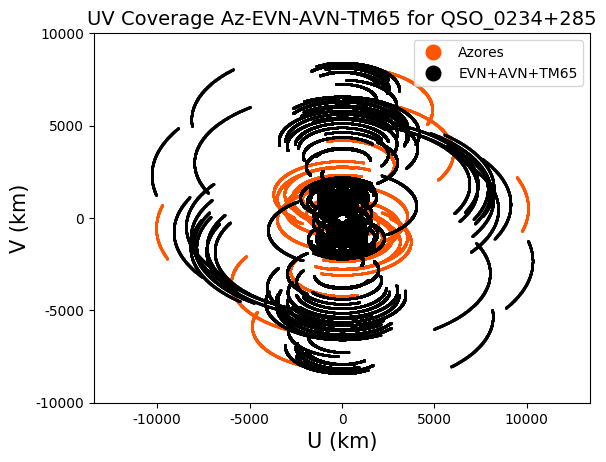}
&
\centering\includegraphics[width=0.475\linewidth]{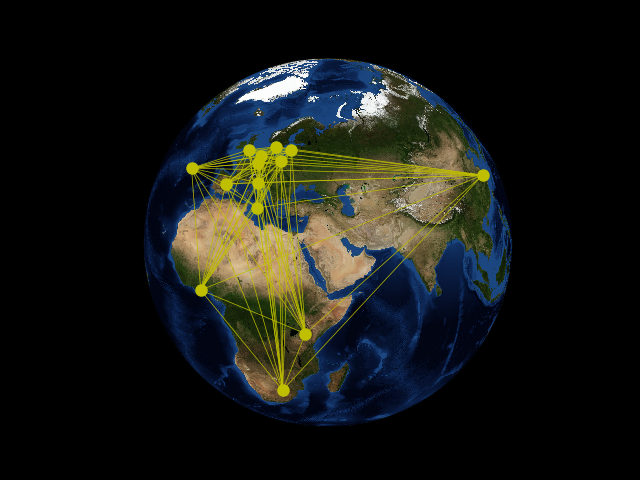}
\end{tabular}
\caption{Left: UV coverage contributions of EVN + AVN + China VLBI network (CVN) Shangai TIANMA65 (China) + Azores (in red). Azores provides unique coverage, and enhances dynamic range of observations of EVN + AVN + Shangai (China) stations added. Right: Baselines of EVN + AVN + China VLBI network (CVN) Shangai TIANMA65 (China) + Azores used in the simulation. Source used: QSO 0234+285.}
\end{figure*}

\begin{figure*}[h]
\begin{tabular}{ll}
\centering\includegraphics[width=0.475\linewidth]{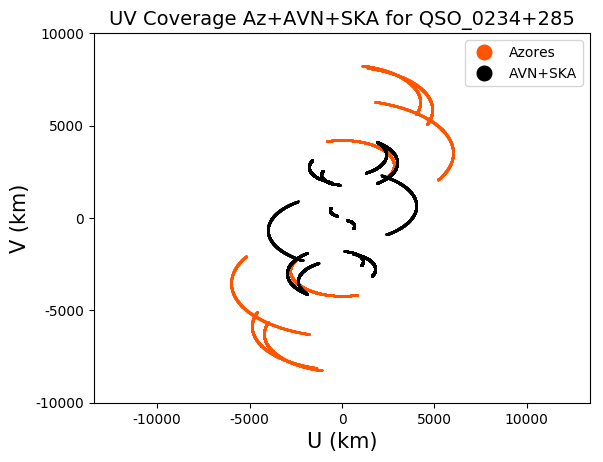}
&
\centering\includegraphics[width=0.475\linewidth]{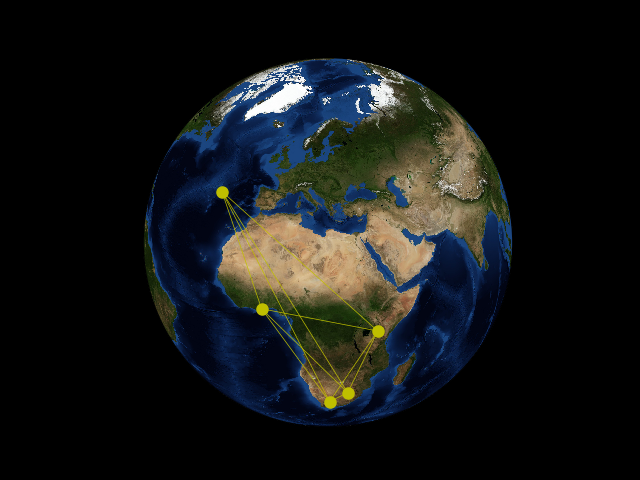}
\end{tabular}
\caption{Left: UV coverage contributions of AVN + SKA-Mid + Azores (in red). Azores provides unique coverage, and enhances dynamic range of observations. Right: Baselines of AVN + SKA-Mid + Azores used in the simulation. Source used: QSO 0234+285.}
\end{figure*}

\begin{figure*}[h]
\begin{tabular}{ll}
\centering\includegraphics[width=0.475\linewidth]{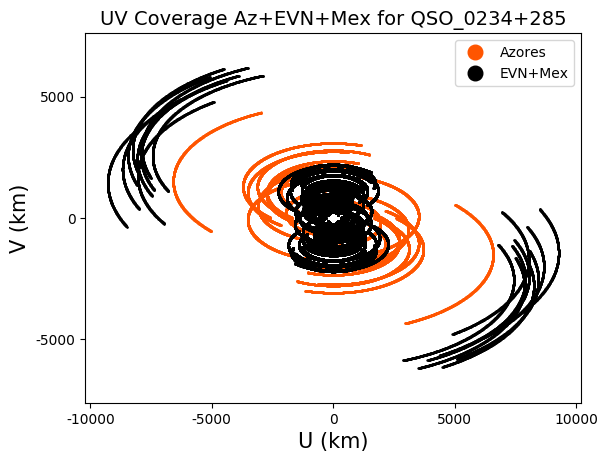}
&
\centering\includegraphics[width=0.475\linewidth]{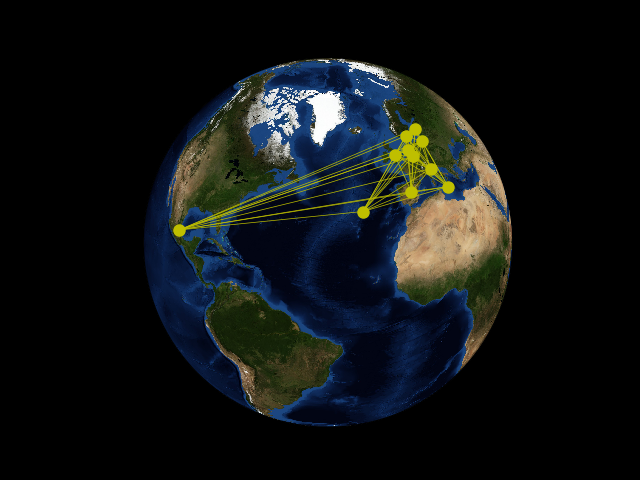}
\end{tabular}
\caption{Left: UV coverage contributions of EVN + Tulancingo2 (Mexico) + Azores (in red). Azores provides unique coverage, and enhances dynamic range of observations. Right: Baselines of EVN + Tulancingo2 (Mexico) + Azores used in the simulation. Source used: QSO 0234+285.}
\end{figure*}

\begin{table*}[h]
\centering
\begin{tabular}{l l c c c c c c}
\hline
\textbf{Station} &\textbf{Code}& \makecell{\textbf{Lat.(deg)}\\ \textbf{Long.(deg)}}&\textbf{\makecell{EVN\\+\color{red}AZ}} & \textbf{\makecell{EVN+AVN\\+\color{red}AZ}} & \textbf{\makecell{EVN+AVN\\+\color{red}AZ \color{black}+TM65}} & \textbf{\makecell{AVN+\color{red}AZ\\ \color{black}+SKA}}
& \textbf{\makecell{EVN+\color{red}AZ\\ \color{black}+Tulancingo2}}\\
\hline
\hline
\makecell[l]{EFLSBERG\\(Germany)}  &Ef/Eb & \makecell[r]{ 50.52485\\   6.88366} & X & X & X & &X\\
\hline
\makecell[l]{MEDICINA\\(Italy)} &Mc  & \makecell[r]{ 44.52048\\  11.64695} & X & X & X& &X\\
\hline
\makecell[l]{NOTO\\(Italy)}  &Nt     & \makecell[r]{ 36.87606\\  14.98903} & X & X & X& &X\\
\hline
\makecell[l]{ONSALA85\\(Sweden)} &On-85  & \makecell[r]{ 57.39307\\  11.91778} & X & X & X& &X\\
\hline
\makecell[l]{YEBES40M\\(Spain)} & Ys & \makecell[r]{ 40.52467\\  -3.08681} & X & X & X& &X\\
\hline
\makecell[l]{TORUN\\(Poland)}  &  Tr  & \makecell[r]{ 53.09547\\  18.56400} & X & X & X& &X\\
\hline
\makecell[l]{WESTERBORK\\(Netherlands)}& Wb& \makecell[r]{ 52.91526\\   6.59396} & X & X & X& &X\\
\hline
\makecell[l]{IRBENE\\(Latvia)}& Ir& \makecell[r]{ 57.55334\\   21.85477} & X & X & X& &X\\
\hline
\makecell[l]{JODRELL BANK\\(UK)}&\makecell[l]{Jb-1\\Lovell76}& \makecell[r]{ 53.23625\\   -2.30715} & X & X & X& &X\\
\hline
\makecell[l]{\color{red}AZORES\\\color{red}(Portugal)} &    & \makecell[r]{ \color{red}37.79071\\ \color{red}-25.66475} & \color{red}X & \color{red}X & \color{red}X & \color{red}X& \color{red}X\\
\hline
\makecell[l]{KUNTUNSE\\(Ghana)}&   & \makecell[r]{  5.75052\\  -0.30517} &   & X & X&X \\
\hline
\makecell[l]{LONGONOT\\(Kenya)} &  & \makecell[r]{ -1.01698\\  36.49688} &   & X & X&X\\
\hline
\makecell[l]{HART\\(South Africa)} & Hh & \makecell[r]{-25.88970\\  27.68541} &   & X & X&X\\
\hline
\makecell[l]{TIANMA65\\(China)}& \makecell[l]{Tm65,\\T6}& \makecell[r]{ 31.09209\\ 121.13597} &   &   & X& \\
\hline
\makecell[l]{TULANCINGO2\\(Mexico)}& & \makecell[r]{ 20.06356\\ -98.43433} &   &   & & &X\\
\hline
\makecell[l]{SKA-Mid\\(South Africa)}& & \makecell[r]{ -30.71280\\ 21.44306} &   &   & &X \\
\hline
\end{tabular}
\label{tab2}
\caption{Summary of the Stations represented on Figures. 3-12. In red the Earth Station in São Miguel (Azores) -- Intelsat Standard A 32m dish (see Figure 2).}
\end{table*}

The Azores is the only place in the North Atlantic that can provide the much necessary East-West baselines. Additionally, the Azores can greatly enhance the recent onset of the Africa VLBI network (AVN), a much necessary scientific "ally" to the EVN and its near-term strategy as a southern extension of the EVN-AVN broader array and contribute to extend the MeerKAT / Square Kilometre Array (SKA) with its midfrequency core in South Africa, see  (eg. Fig. 6). The Azores antenna location also enables observation of sources below the celestial equator up to $\sim$ -52$^{\circ}$ and therefore contributes to the exploration of Southern sky deep fields and sources explored by ESO facilities, ALMA and SKA. A more detailed analysis of an observational strategy considering different source declinations shall be performed in a follow up study using the methods outlined in \citet{2020arXiv200308642A,2020arXiv200104576G,2011Heywood,2011arXiv1103.3600K}.

\subsection{Networking, Timing and e-Services for e-VLBI}

As it is expected by VLBI networks like the EVN and AVN, large radio telescopes will regularly participate in VLBI and electronic VLBI (e-VLBI) observations, adding value to African and Eurasia e-networks. This requires good quality of service delivery for data transfer. The Azores Intelsat Standard-A antenna station compound in S. Miguel has besides the 32-metre station, a 12-metre station for SATCOM communication link with the Azores Western group (Flores and Corvo Islands). The compound is served by a dedicated 1 Gbps point of presence (Giga-PoP) link connecting to the Azores fiber loop at 10 Gpbs. The Azores itself have very good connectivity having access to the new submarine cables nearby and the archipelago is connected to mainland Portugal. This should enable e-VLBI operations with 1 Gbps speed (i.e. 16 channels each with 16 MHz bandwidth and 2 bit resolution). In fact, the RAEGE station in St. Maria already participates in e-VLBI experiments with interferometric fringe analysis for geodetic purposes. We note that operators and NRENS are already implementing 100 Gbit/sec enabling large data streams between Africa or Australia and Europe thus opening great prospects for added capacity and real time correlations.

Precise Timing is necessary to enable phase coherency when correlating signals. A maximum permitted coherence loss of 2\% equates to 0.2 radians of phase error, which, at a maximum observing frequency of 6 GHz, corresponds to accuracies of $\sim{10}$ ps. Also, high precision long-term timing is necessary for astrophysical
phenomena such as pulsars and transients, nowadays major scientific cases. In particular, pulsar monitoring experiments require timing accuracies of 10 ns over time periods of 10 years. Overall, synchronization to an absolute time provider is required for system management, antenna pointing, beam steering, time stamping of data and producing regular timing ticks. From the experimental point of view, the clocks should enable a frequency standard for the Local Oscillators (LOs), digitizer clocks and other devices.

Timing and synchronization to the required stability should be provided by a local hydrogen maser (in the VLBI case) or from a distributed frequency reference signal, locked to a central ‘master’ hydrogen maser frequency standard. Since Azores islands are connected by their Azorean fiber ring loop, the SATCOM antenna could also be synchronized after time transfer over fiber from the operating “clock master” at St. Maria RAEGE station with White Rabbit technologies, a framework for Ethernet-based network for general purpose data transfer and
sub-nanosecond accuracy time transfer being used in radio astronomy and space, for instance by EVN \citep{2009Serrano,2017Kaur}. This is also the case for the SKA1, where White Rabbit is being considered as a prime option for accurate time transfer across the SKA1 network \citep{2017Grainge}. As a spin-off, recently it has been shown recent optical fiber submarine cable can be explored as deployed seismographs, therefore also enhancing the scientific instruments infrastructure in the Azores. Indeed, the reinforcement of the Atlantic radio astronomical and space cluster favors the emergence of new testbeds on time distribution technologies across digital networks.

\section{Science goals}
Scientifically, VLBI enables imaging with high angular resolution of bright compact radio sources, a crucial aspect for multi-messenger astronomy. We highlight in the following discussion a few of the many potential science cases that a VLBI network can contribute. We point out it is beyond the scope of this work to describe these in details and we refer to the EVN Science Vision \citep{2020venturi} that covers the following broad areas: cosmology, galaxy formation and evolution, innermost regions of active galactic nuclei, explosive phenomena and transients, stars and stellar masers in the Milky Way, celestial reference frames and space applications. Besides the cooperation with other radio facilities, the EVN vision also draws a close discussion on synergies with multi-band/multi-messenger instruments and provides roadmap recommendations. Further focused discussions on the SKA VLBI synergies are developed in \citet{2019arXiv190308627G} and references there in.  We can cite as most important science cases: 

\begin{itemize}
\item Quasars – physics of relativistic jets, calibrators, astrometry (also in the framework of Gaia mission)
\item Microquasars – behavior and parallaxes – distances
\item Pulsars – proper motions and parallax, interstellar scattering, emission region size
\item Transient sources, including Novae, Fast Radio Bursts (FRBs) and Gravitational Waves (GW) triggered events
\item Neutrino science
\item Supernovae – behavior of exploding star remnants
\item Interacting binary star behavior
\item E-VLBI and TOO VLBI through internet – rapid response to new events (triggered by Gravitational waves or cosmic rays’ events).
\item Near-Earth object tracking capability gained by the combined optical and VLBI methods
\end{itemize}

Additionally, VLBI capable telescopes can also operate in “single dish” mode, since e-VLBI experiments run only for some weeks in a row. An Azores SATCOM antenna has the advantage of being closer to the Equator than most of other EVN stations (see Table 2). Therefore, it has access to a larger area of the sky and can observe a much bigger fraction of the Milky Way compared with EVN radio telescopes based in the Northern Hemisphere. The high sensitivity of such a large dish can be used for these “single dish” opportunities:

\begin{itemize}
\item Spectroscopy with narrowband multi-channel receiver
\item Monitor masers in star-forming regions eg. for periodic variations (methanol at 6668 MHz)
\item Survey formaldehyde absorption in Milky Way dark clouds (4829 MHz)
\item Pulsar observing with wideband multi-channel receiver
\item Monitor pulsars for glitches, long term behavior, proper motion
\item Search / monitor for intermittent pulsars and transients (Rotating RAdio Trasients -- RRATs)
\item Radio continuum flux measurement with wideband multi-channel receiver
\item Monitoring of Gamma-ray flare sources
\item Planetary sciences and Space Situational Awareness (SSA) or planetary defence: Single dish observations of NEOs or comets at OH maser radiation line.
\end{itemize}

On this last topic Space Situational Awareness (SSA) programs became most important for space agencies when considering the security of space assets, in particular for the emergent satellite constellations: Space Weather phenomena, Near-Earth Objects and Space debris. Of particular interests for the SSA is the European Space Surveillance and Tracking (SST) segment program fostering the creation of a network infrastructure of new sensors (optical, radio) capable of monitoring accurately the space debris field affecting commercially interesting orbits. Radiotelescopes and radio interferometers are examples of ground based passive radio sensors capable of providing important ancillary space debris and NEOs information if they are set in the appropriate configuration required for this.

There have been considerable interests in promoting radar sensors in Portugal mainland and the Azores (in particular at its Eastern group) since both Portugal mainland or the Azores could operate providing an interesting connecting between sensors like the Haystack Auxiliary Radar, operated by the Massachusetts Institute of Technology Lincoln Laboratory (MIT LL) for the NASA Orbital Debris Program Office (ODPO) and the European sensors. 

Larger antennas could also be used for piggy-back space debris or NEOs monitoring up to GEO orbits using bistatic or multistatic surveying configurations. In 2001 the Medicina radiotelescope was used together with the Goldstone (USA) and the Evpatoria (Ukraine) transmitters to perform planetary radar observations of Near-Earth Asteroid 1998 WT24 \citep{2004P&SS...52..325D}.
Therefore, the large S. Miguel SATCOM dish could also contribute in the future if appropriately configured with a piggy back receiver for observations or monitoring of space debris up to geostationary orbits, which due to distance are more difficult to observe, after receiving echos of a faraway transmitter in a bistatic configuration. The bistatic radar capabilities of the VLBI radiotelescopes for space debris detection and tracking is unsurpassed in combination with facilities in other radio bands. 

\subsection{Astrophysical Transients}
Among the several science cases previously listed, the discovery of new astrophysical transients will require detailed follow up (VLBI and single dish modes) and monitoring (single dish mode), in particular in a multi-band/multi-messenger era.  In follow up mode, we will require to be triggered by other observatories such as the gravitational wave (LIGO/Virgo and eventually LISA), x-ray satellites, CTA, HESS; ICECUBE (neutrino), gamma-rays telescopes (Fermi), SKA among other instruments. Besides, Azores will also contribute to the transient sky studies since it will host a node of the SVOM space mission ground alert network (see Table 1) that is dedicated to the detection and discovery of GRBs and therefore it is expected a particular attention to be provided to this important scientific topic.

Of major interest is the follow up and monitoring of the merger of binary neutron star mergers which lead to gravitational waves \citep{PhysRevLett.119.161101} and a kilonova explosion observed in the electromagnetic spectrum from gamma-rays to radio waves \citep[e.g.,][]{2017ApJ...848L..13A,2017ApJ...848L..16S,2017ApJ...848L..18N,2017ApJ...848L..19C,2017ApJ...848L..20M,2017ApJ...848L..21A,2017ApJ...848L..22B}. These explosive events produce the so called r-process elements which are elements heavier than iron. High angular resolution of these explosions will permit us to determine how these elements are distributed throughout the interstellar medium. 

Interesting prospects that the Azores antenna provides in terms of resolution is on the field of thermonuclear, nova, eruption which occurs on the surface of a white dwarf star \citep[see, e.g.,][]{2014ASPC..490.....W}. Here considerable amount of material is expelled into the interstellar medium and reach peak flux densities as high as a few 10s of milli-Jy (at 1.4 GHz). The radio light-curves originally were modelled as free-free thermal emission from a expanding photosphere, which permits us to determine the ejected mass, expansion velocities and kinetic energy of the eruptions \citep{1979AJ.....84.1619H,2015aska.confE..62O}. However, over the last decade the radio light-curve has been characterised by synchrotron emission during the early phase \citep[within the first few weeks, e.g.,][]{2009MNRAS.395.1533E}.

There were two scenarios to explain the origin of synchrotron emission. 1) Interaction of the ejecta with pre-existing circumstellar material, or 2) inter-ejecta shocks (\citep{2014Natur.514..339C}. Of particular relevance for VLBI, and as demonstrated in \citet{2014Natur.514..339C}, are the EVN 5GHz observations which were able to resolve the location where the synchrotron emission arose from. From the observations, we were able to infer that gamma-ray emission observed in these novae must arise from shocks between slow and fast winds. However, in order to test this scenario, we require a large sample of novae at the highest resolution, only afforded by VLBI observations. Furthermore, VLBI observations will allow us to  understand why we only observe gamma-ray emission in some novae and not in others. See the EVN Science vision and its chapter on astrophysical transients for further details \citep{2020venturi}.

\section{SKA and VLBI}
\label{S:2}

The Atlantic scientific dynamics will greatly accelerate with the deployments of the SKA and the Africa VLBI Network (AVN). Since the SKA 1 (MID) will behave as a big and very sensitive station, the expansion of its VLBI capabilities (SKA-VLBI) with addition of stations across Africa and eventually the EVN will greatly broaden the science of the SKA, see \citet{2015Paragi} for an extensive analysis. In fact, the VLBI will be much necessary to increase substantially spatial resolution. For the future Phase 2, it is expected that the outer SKA stations will greatly enhance the VLBI capabilities through a distributed baseline configuration of up to a few thousands of km, eventually merging it with the existing single sited VLBI station networks.

This high spatial resolution capability has long been considered an essential part of the SKA \citep{2000Garrett,2004Gurvits,2004Fomalont,2008Schilizzi,2012Godfrey}. The science goals are best achieved with SKA1 by forming phased-array elements from SKA1-MID observing together with existing VLBI arrays in the 1-13.5 GHz frequency, \citep{2019arXiv190110361P,2019arXiv190308627G}.

\begin{table*}[h]
\centering
\begin{tabular}{l l l l}
\hline
\textbf{SWG} & \textbf{Science Objectives} & \textbf{\makecell{SWG\\Rank}} &\textbf{\makecell{VLBI\\with:}}\\
\hline
\hline
Pulsars & \makecell[l]{High precision timing for testing Gravity and\\GW detection} & 1/3 & \makecell[l]{LOW/\\MID}\\
\hline
HI & \makecell[l]{Resolved HI kinematics and morphology of\\$\sim{10^{10}}$ M$_\odot$ galaxies out to z$\sim0.8$} & 1/5 & \makecell[l]{LOW/\\MID}\\
\hline
Transients & \makecell[l]{Solve missing baryon problem at z$\sim{2}$ and\\determine Dark Energy Equation of State} & 1/4 & MID \\
\hline
\makecell[l]{Cradle\\of life} & \makecell[l]{Map dust grain growth in the terrestrial planet\\forming zones at a distance of 100pc} & 1/5 & MID \\
\hline
Continuum&\makecell[l]{Star Formation history of the Universe (SFHU)\\I+II. Non-Thermal and Thermal processes}&\makecell[l]{1+2\\/8}&MID \\
\hline
\end{tabular}
\caption{Summary of important SKA High Priority Science Objectives that will be impacted by a VLBI observational mode. For further references see \citep{2019arXiv190308627G,2019arXiv190110361P} and references therein.}
\end{table*}

Furthermore, VLBI astrometry will remain a very important tool for astrophysics enhancing and surpassing Gaia astrometry products, with several applications to navigation and Reference Systems. For example, pulsar parallax measurements using SKA-VLBI will play an essential role in several high impact areas, including strong field tests of gravity in relativistic binary systems, tomographic mapping of the Galactic magnetic field and mapping the ionized interstellar plasma in the Galaxy, and the physics of neutron stars, as well as detecting the gravitational wave background.

SKA-VLBI will provide very sensitive, milliarcsecond (mas) resolution imaging that is important for the study of the physics of jets in Active Galactic Nuclei \citep[AGNs;][]{2015Agudo}, the formation of the very first generation of Super Massive Black Holes (SMBHs) in the Universe and in connection with the Cherenkov Telescope Array (CTA) or the Pierre Auger it will help reveal the nature of the large population unidentified high-energy sources \citep{2015Giroletti}.

Since the VLBI configuration with SKA will be rather conventional and similar to other tied beam arrays, an addition of a station to cover the Atlantic gap may greatly enhance the scientific return of SKA-VLBI. As our analysis shows in chapter 3, the Azores antenna contributes to improve both the sensitivity and angular resolution of EVN+AVN arrays and SKA.

\section{Conclusions}

Azores has a unique location across the Atlantic for space and radio astronomy infrastructure of world-class capabilities: its location covers the so called “Atlantic gap”. An Astronomical VLBI capability in Azores, via conversion of former 32-metre SATCOM antenna in S. Miguel or the installation of a couple of MeerKAT/SKA1-MID dishes would offer long and unique East-West visibilities to SKA and existing VLBI networks (AVN and EVN). It enhances combined array resolution up to $\sim$ 1.3 miliarcsecond in radio wavelengths and would contribute to fill the overall array image UV-plane resulting in greater image fidelity. Additionally, this facility would contribute to enhance the southern extension of the EVN-AVN broader array and to bridge to the MeerKAT / Square Kilometre Array (SKA) Phase 1 mid frequency core in South Africa.  This is an opportunity for cutting-edge research by developing a world-class infrastructure for radio astronomy and space exploration at a relatively modest investment and great prospects of sustainability. 

Radio astronomy techniques enable ground-breaking studies of the widest variety of cosmic phenomena: VLBI is the only technique enabling the imaging of the horizon of Black Holes amplifying the impact of larger projects like SKA; it enables unsurpassed precision in the tracking and reception of planetary probes in the deep Solar system beyond Mars. On itself, the SKA-VLBI mode will significantly improve the connection between the celestial reference frames defined in the optical and radio bands and will have a profound effect on a large number of fields within astronomy. The latitude of the Azores antenna also makes Southern sky regions accessible up to -52$^{\circ}$ declination thus enabling observations from joint research programs or follow up of Key Science Programs or Legacy programs from ESO, including ALMA and VISTA, SKA and other facilities targeting transient sources or nuclear explosions in the Universe like Auger, HESS and CTA.

The extension of VLBI services to SKA will have the following impact:

\begin{itemize}
\item Continued use of very expensive installation that is now or is becoming redundant, at relatively lower cost, with insertion in international networks with access to funding programs.
\item Reinforcement of links to global radio astronomy networks through VLBI and international usage of a national facility.
\item In country training in practical radio astronomy with synergies/aligned with RAEGE/AIR goals:
\begin{itemize}
    \item Single-dish research
    \item Very Long Baseline Interferometry (VLBI) for high angular resolution imaging and space
\end{itemize}
\item Create a pool of astronomers and experimentalists able to use current and future large scale radio telescope arrays
\item Opportunities for training, research and development in engineering and technology
\begin{itemize}
    \item low-noise microwave feeds and receivers
    \item analogue and digital electronics
    \item digital signal processing, Big Data software engineering
\end{itemize}
\item Stimulate interest in science, engineering and technology through outreach program connected to very high impact science.
\item Spinoffs: reinforcement of Azores digital infrastructures and contribute to SSA/SST activities.
\item Promote the development of the National Space Agency (PTSPACE) activities and the highly synergistic interactions with/within the Atlantic International Research Centre (AIR).
\end{itemize}

\section*{Acknowledgements}
We warmly thank Francisco Colomer (JIVE - Joint Institute for VLBI ERIC) and Leonid Gurvits (JIVE and Astrodynamics and Space Missions, TU Delft) for the full and comprehensive support about the international interest on the Azores VLBI cluster. The team warmly acknowledges Altice Portugal, C\'esar Malheiro and Joaquim Diniz for all the kind information, access to the SATCOM compound in S. Miguel and further collaboration in Azores, Lu\'{\i}s Santos from Estrutura de Miss\~ao dos A\c cores para o Espa\c co of Azores Government for all the kind collaboration and local support. The team acknowledges financial support from the Aga Khan Development Network and the Fundação para a Ciência e a Technologia, Portugal, for the Science and Technology Cooperation DOPPLER - Development of PaloP knowLEdge in Radioastronomy, project number 333197717. V.A.R.M.R. acknowledges financial support from the Funda\c c\~ao para a Ciência e Tecnologia (FCT) in the form of an exploratory project of reference IF/00498/2015, funded by Programa Operacional Competitividade e Internacionaliza\c c\~ao (COMPETE 2020) and FCT, Portugal. We acknowledge support from ENGAGE SKA (POCI-01-0145-FEDER-022217), and PHOBOS (POCI-01-0145-FEDER-029932), funded by COMPETE 2020 and FCT, Portugal. A.C. acknowledges support from CFisUC projects UIDB/04564/2020 and UIDP/04564/2020.This work was also funded by FCT and Ministério da Ciência, Tecnologia e Ensino Superior (MCTES) through national funds and when applicable co-funded EU funds under the project UIDB/50008/2020-UIDP/50008/2020 and UID/EEA/50008/2019. S.A. acknowledges financial support from the Centre for Research and Development in Mathematics and Applications (CIDMA) strategic project UID/MAT/04106/2019.  We also warmly thank the comments and suggestions of the anonymous referees that much improved this manuscript.

\bibliographystyle{model5-names}
\biboptions{authoryear}

\bibliographystyle{model5-names}
\biboptions{authoryear}

\end{document}